\documentclass[a4paper,11pt]{article}
\usepackage{jcappub} % for details on the use of the package, please see the JINST-author-manual

\usepackage{lineno}
\usepackage{xcolor}
\usepackage{multirow}
\usepackage{hyperref}
\usepackage{caption}
\usepackage{subcaption}
\usepackage{floatrow}
\usepackage{bbm}
\newfloatcommand{capbtabbox}{table}[][\FBwidth]

\title{Impact of cosmology dependence of baryonic feedback in weak lensing}

\author[a]{Pranjal R. S.,}
\author[a,b]{Elisabeth Krause,}
\author[c,d]{Klaus Dolag,}
\author[e]{Karim Benabed,}
\author[a,b]{Tim Eifler,}
\author[e]{Emma Ayçoberry,}
\author[e]{and Yohan Dubois}
\affiliation[a]{Department of Astronomy/Steward Observatory, University of Arizona, 933 North Cherry Avenue, Tucson, AZ 85721, USA}
\affiliation[b]{Department of Physics, University of Arizona, 1118 E Fourth Street, Tucson, AZ 85721, USA}
\affiliation[c]{Universitäts-Sternwarte, Fakultät für Physik, Ludwig-Maximilians-Universität München, Scheinerstr. 1, D-81679 München, Germany}
\affiliation[d]{Max Planck Institute for Astrophysics, Karl-Schwarzschild-Str. 1, D-85741 Garching, Germany}
\affiliation[e]{Institut d'Astrophysique de Paris, UMR 7095, CNRS \& Sorbonne Universit\'e, 98 bis boulevard Arago, F-75014 Paris, France}

\emailAdd{pranjalrs@arizona.edu}

\abstract{Robust modeling of non-linear scales is critical for accurate cosmological inference in Stage IV surveys. For weak lensing analyses in particular, a key challenge arises from the incomplete understanding of how non-gravitational processes, such as supernovae and active galactic nuclei — collectively known as \textit{baryonic feedback} — affect the matter distribution. Several existing methods for modeling baryonic feedback treat it independently from the underlying cosmology, an assumption which has been found to be inaccurate by hydrodynamical simulations. In this work, we examine the impact of this coupling between baryonic feedback and cosmology on parameter inference at LSST Y1 precision. We build mock 3$\times$2pt data vectors using the \textit{Magneticum} suite of hydrodynamical simulations, which span a wide range of cosmologies while keeping subgrid parameters fixed. We perform simulated likelihood analyses for two baryon mitigation techniques: (i) the Principal Component Analysis (PCA) method which identifies eigenmodes for capturing the effect baryonic feedback on the data vector and (ii) \textsc{HMCode2020} \citep{Mead_2021} which analytically models the modification in the matter distribution using a halo model approach. Our results show that the PCA method is robust to the coupling between cosmology and baryonic feedback, whereas, when using \textsc{HMCode2020} there can be up to $0.5\sigma$ bias in $\Omega_\text{m}$–$S_8$. For \textsc{HMCode2020}, the bias also correlates with the input cosmology while for PCA we find no such correlation.}

\begin{document}
\maketitle
\flushbottom

\section{Introduction}
\label{sec:intro}
The concordance $\Lambda$CDM model has been rigorously tested by the current generation of wide-field surveys and has been remarkably successful in explaining a wide range of observations \citep[e.g.][]{Planck2020, KiDS_cosmic_shear,Alam2021,Abbott2022, Brout2022, Miyatake2023, Sugiyama2023, DESI_Y1, Qu2024}. However, challenges to the model have also emerged and among the longstanding issues is the discrepancy between the amplitude of matter fluctuations measured at lower redshifts and those predicted by cosmic microwave background (CMB) experiments \citep[e.g.][]{Valentino2021}. Upcoming Stage IV weak lensing surveys, such as the Rubin Observatory’s Legacy Survey of Space and Time (LSST\footnote{\href{https://www.lsst.org}{\nolinkurl{https://www.lsst.org}}}, \citep{LSST}), \textit{Nancy Grace Roman Space Telescope} (\textit{Roman}\footnote{\href{https://roman.gsfc.nasa.gov}{\nolinkurl{https://roman.gsfc.nasa.gov}}}, \citep{Roman}), and \textit{Euclid}\footnote{\href{https://sci.esa.int/web/euclid}
{\nolinkurl{https://sci.esa.int/web/euclid}}} \citep{Euclid}, will play a crucial role in addressing these issues and exploring extensions to the $\Lambda$CDM paradigm.

While non-linear scales probed by weak lensing offer a rich source of information, these scales are subject to a complex interplay between gravitational evolution and baryonic physics. The formation of stars and supermassive black holes in galaxies triggers events such as supernovae and active galactic nuclei (AGN), which inject energy into the surrounding medium. These processes regulate star formation by preventing the collapse of cold gas and expelling it into the circumgalactic  and intracluster medium. The redistribution of gas, in turn, induces a gravitational back-reaction on the dark matter halo, further altering the matter distribution. State-of-the-art hydrodynamical simulations have demonstrated that feedback mechanisms can suppress the matter power spectrum by $\mathcal{O}(10\%)$  at 1 Mpc$^{-1}h<k<10\, \text{Mpc}^{-1}h$.

Several methods have been developed to address the effects of baryonic feedback in weak lensing analyses. One common strategy is to exclude the range of scales that are not modeled with sufficient accuracy and may thus bias the inferred cosmology \citep[e.g.][]{Krause2017,Li2023}. Another approach is to utilize analytic prescriptions to capture the modifications in dark matter halos through a set of baryonic parameters, these parameters can be calibrated from hydrodynamical simulations and/or varied alongside cosmology during the inference process. Examples of this approach include, HMCode which uses the halo model formalism to modify halo shapes and their baryonic content \citep{Mead_2015, Mead_2021}, and the  \textit{baryonification} method which modifies particle positions in gravity-only simulations to emulate the impact of baryonic physics \citep{Schneider2015, Arico2021}. Alternatively, the PCA method directly models the impact of baryonic feedback at the level of summary statistics by employing a set of eigenmodes and marginalizing over their amplitudes \citep{Eifler2015, Huang_19}.

State-of-the-art baryon mitigation techniques generally model the growth of structure and baryonic feedback independently despite the intrinsic link between galaxy formation and the underlying cosmology. Previous studies using hydrodynamical simulations have found that, for a given subgrid physics implementation, the suppression in the matter power spectrum varies with cosmological parameters \citep[e.g.][]{Schneider2020, Stafford2020, Vandaalen2020, Arico2021, Delgado2023}. Feedback effects are more pronounced in cosmologies where halos have a shallower gravitational potential well (e.g. due to changes in concentration) or higher baryon fractions \citep{Elbers2024}. While there is evidence that this coupling has negligible impact on cosmological constraints even for a \textit{Euclid}-like survey \citep{Parimbelli2019,Schneider2020}, these studies have been limited by the accuracy of the analytical fitting functions used for building mock data. Therefore, it is important to further investigate these effects in the context of Stage IV surveys which will push us deeper into the non-linear regime \citep{Chisari2019}. Note that the dependence of baryonic feedback on cosmology is distinct from the degeneracy between the two, the latter simply implies that their effects on observables are indistinguishable.

The goal of this work is to ascertain whether baryon mitigation techniques, which assume baryonic feedback to be independent of cosmology, can deliver unbiased cosmological constraints. In particular, we consider two such techniques: PCA  \citep{Eifler2015, Huang_19} and the \textsc{HMCode2020} model  \citep{Mead_2015, Mead_2021}. The PCA method uses sets of hydrodynamical simulations to identify eigenmodes, also called principal components (PCs), which can be used to model the corrections due to baryons. These PCs are added to the dark matter-only data vector with their amplitude as free parameters. In contrast, the \textsc{HMCode2020} approach explicitly models the modification to the matter power spectrum through a halo model formalism with free parameters for baryonic feedback that are calibrated from simulations. Both these methods have been used for baryon mitigation in weak lensing analyses \citep[e.g.][]{Huang_21, Xu2023, KiDS_cosmic_shear, DESY3_KIDS}.

We characterize the performance of these methods using hydrodynamical simulations covering wide range of fiducial cosmologies. We build mock 3$\times$2pt data vectors using the \textit{Magneticum} hydrodynamical simulation suite, which spans fifteen cosmologies. Using these mock data, we perform simulated likelihood analyses for a LSST Y1-like survey and quantify the parameter bias that arises from the coupling between cosmology and baryonic feedback.

This paper is organized as follows. In section \ref{sec:sims} we describe the hydrodynamical simulations and their basic properties. In section \ref{sec:Pk_suppression} we discuss the suppression in the matter power spectrum measured from the \textit{Magneticum} simulations. Section \ref{sec:analysis} details the theory model, analysis choices and survey design. We present the results in section \ref{sec:results} and analyze the performance of the baryon mitigation techniques across cosmologies. We conclude in section \ref{sec:conclusions}.

\section{Simulations}
\label{sec:sims}

\begin{table}
    \centering
\begin{tabular}{l c c c  c c}
\hline
\hline 
& \(\Omega_\text{m}\) & \(\Omega_\text{b}\) & \(\sigma_8\) & \(h\) &  \(\Omega_\text{b}/\Omega_\text{m}\)\\
\hline C1 & 0.153 & 0.0408 & 0.614 & 0.666  & 0.267\\
C2 & 0.189 & 0.0455 & 0.697 & 0.703  & 0.241\\
C3 & 0.200 & 0.0415 & 0.850 & 0.730  & 0.208\\
C4 & 0.204 & 0.0437 & 0.739 & 0.689  & 0.214\\
C5 & 0.222 & 0.0421 & 0.793 & 0.676  & 0.190\\
C6 & 0.232 & 0.0413  & 0.687 & 0.670  & 0.178\\
C7 & 0.268 & 0.0449  & 0.721 & 0.699  & 0.168\\
C8 (WMAP7) & 0.272 & 0.0456 & 0.809 & 0.704  & 0.168\\
C9 & 0.301 & 0.0460 & 0.824 & 0.707  & 0.153\\
C10 & 0.304 & 0.0504 & 0.886 & 0.740  & 0.166\\
C11 & 0.342 & 0.0462 & 0.834 & 0.708  & 0.135\\
C12 & 0.363 & 0.0490 & 0.884 & 0.729  & 0.135\\
C13 & 0.400 & 0.0485 & 0.650 & 0.675  & 0.121\\
C14 & 0.406 & 0.0466 & 0.867 & 0.712  & 0.115\\
C15 & 0.428 & 0.0492 & 0.830 & 0.732 & 0.115\\
\hline
\hline 
\end{tabular}
\caption{Cosmological parameter combinations for the 15 simulations in the \textit{Magneticum} suite.}
    \label{tab:box3_cosmo}
\end{table}
\subsection{Magneticum}
\label{sec:sim_magneticum}
The \textit{Magneticum}\footnote{\href{ http://www.magneticum.org}{\nolinkurl{ http://www.magneticum.org}}} suite of hydrodynamical simulations \citep{Hirschmann2014, Teklu2015, Dolag2016, Steinborn2016, Bocquet2016, Remus2017, Castro2020} were run using the smoothed particle hydrodynamics code \textsc{P-GADGET3} \citep{Springel2005}. The subgrid physics implementation in these simulations includes radiative cooling, heating, ultraviolet background, star formation \citep{Springel2005b}; stellar evolution, chemical enrichment, and metallicity dependent cooling \citep{Tornatore2007,Wiersma2009}; feedback from supernovae driven galactic winds and active galactic nuclei \citep{Springel2003, Hirschmann2014}. 

Several studies have shown that the \textit{Magneticum} simulations reproduce a wide range of observations spanning from galactic to cluster scales including: the galaxy stellar mass function  \citep{Lustig2022}; AGN luminosity function \citep{Biffi2018}; properties of the intra-cluster medium (e.g. temperature profiles) \citep{McDonal2014, Gupta2017} and the Sunyaev-Zeldovich power spectrum \citep{Dolag2016}. Of particular relevance to this work is the baryon fraction in galaxy groups and clusters, a strong indicator of the  strength of feedback \citep{Vandaalen2020, Salcido2023}, which has also been found to be consistent with observations \citep{Angelinelli2022, Angelinelli2023}. The level of suppression in the matter power spectrum is similar to the BAHAMAS simulations \citep{bahamas}.

In this work we use the multi-cosmology \texttt{Box3/hr} simulations that have a box size of 128 Mpc $h^{-1}$. The suite consists of 15 simulations C1, C2$\dots$C15, numbered in order of increasing $\Omega_\text{m}$. Each simulation is run using a total of $2\times576^3$ particles with gas and dark matter mass resolutions of $m_{\text{gas}}=1.4\times10^8$ M$_\odot h^{-1}$ and $m_{\text{DM}}=6.9\times10^{8}$ M$_\odot h^{-1}$, respectively. The cosmological parameter combinations for the \textit{Mangeticum} simulations are listed in table \ref{tab:box3_cosmo}.

Out of these simulations, only C8, which is run at the fiducial \textit{WMAP}7 cosmology $\{\Omega_\text{m}, \Omega_\text{b}, \sigma_8, h, n_\text{s}\}=\{0.272, 0.0456, 0.809, 0.704, 0.963\}$ \citep{Komatsu2011}, is calibrated to reproduce observations. The remaining simulations adopt the same subgrid parameters, which ensures that any variations in feedback strength can be attributed solely to changes in cosmology.

\begin{table*}
    \centering
\begin{tabular}{ l r r r  r r}
\hline
\hline 
Simulation & Box Size \hspace{1.5pt}  & Total Particles & Dark matter & Initial gas \hspace{3pt} & Cosmology \\
&  & & particle mass &  particle mass &\\
&(Mpc $h^{-1}$) & &(M$_\odot\,h^{-1}$) &(M$_\odot\,h^{-1}$) &\\
\hline 
Eagle                     & 67.77 & $2\times1504^3$ & 6.57 $\times10^6$& 1.23 $\times10^6$          & \textit{Planck}2013\\
IllustrisTNG                    & 75 & $2\times1820^3$ & 5.06 $\times 10^6$ & 9.44 $\times 10^5$ & \textit{Planck}2013 \\
Illustris                 &  75 & $2\times1820^3$ & 4.41 $\times10^6$& 8.87 $\times10^5$             & \textit{WMAP}7 \\
MassiveBlack-II           & 100 & $2\times1972^3$ & 1.1 $\times10^7$& 2.2 $\times10^6$              & \textit{WMAP}7 \\
Horizon-AGN               & 100 & $2\times1024^3$ & 5.6 $\times10^7$& 7 $\times10^6$                & \textit{WMAP}7 \\
cosmo-OWLS &  400 & $2\times1024^3$ & 3.75 $\times10^9$& 7.54 $\times10^8$                           & \textit{WMAP}7\\
BAHAMAS   & 400 & $2\times1024^3$ & 3.85 $\times10^9$& 7.66 $\times10^8$                             & \textit{WMAP}9 \\
\hline
\hline 
\end{tabular}
    \caption{Simulations used for PCA training.}
    \label{tab:sim_pca}
\end{table*}

\subsection{Simulations for PCA training}
\label{sec:sim_PCA}
We use an independent set of simulations for PCA training: Eagle \citep{eagle}, Illustris \citep{Illustris}, IllustrisTNG \citep{TNG}, MassiveBlack-II \citep{mb2},  HorizonAGN \citep{horizonAGN}, cosmo-OWLS \citep{cOWLS}, and BAHAMAS \citep{bahamas}. The latter two simulations include additional scenarios varying the strength of AGN feedback using the $\Delta T_{\mathrm{heat}}$ parameter, which determines the amount of energy the black hole deposits to the neighboring gas particles. For cosmo-OWLS and BAHAMAS $\Delta T_{\mathrm{heat}}= (10^{8.0}, 10^{8.5}, 10^{8.7})$ K and  $(10^{7.6},10^{7.8}, 10^{8.0})$ K, respectively. The simulation characteristics are summarized in table \ref{tab:sim_pca}.

These simulations cover a broad variety of subgrid physics implementations, e.g. Massive Black-II has relatively weak AGN feedback efficiency which results in an over-prediction of the abundance of massive galaxies at low redshifts \citep{mb2}, the violent radio mode AGN feedback in Illustris causes massive halos to be almost devoid of gas \citep{Genel2014}, and the AGN feedback in the Eagle simulation only injects energy through a thermal channel rather than the commonly used quasar- and radio-mode. As a result, the power spectrum suppression at $z=0$ can vary up to 30\% across these simulations along with different evolution with redshift -- these variations ensure that the eigenmodes identified by the PCA have sufficient flexibility.

\section{Impact of cosmology on baryonic feedback}
\label{sec:Pk_suppression}

\begin{figure*}
    \centering
    \includegraphics[width=\linewidth]{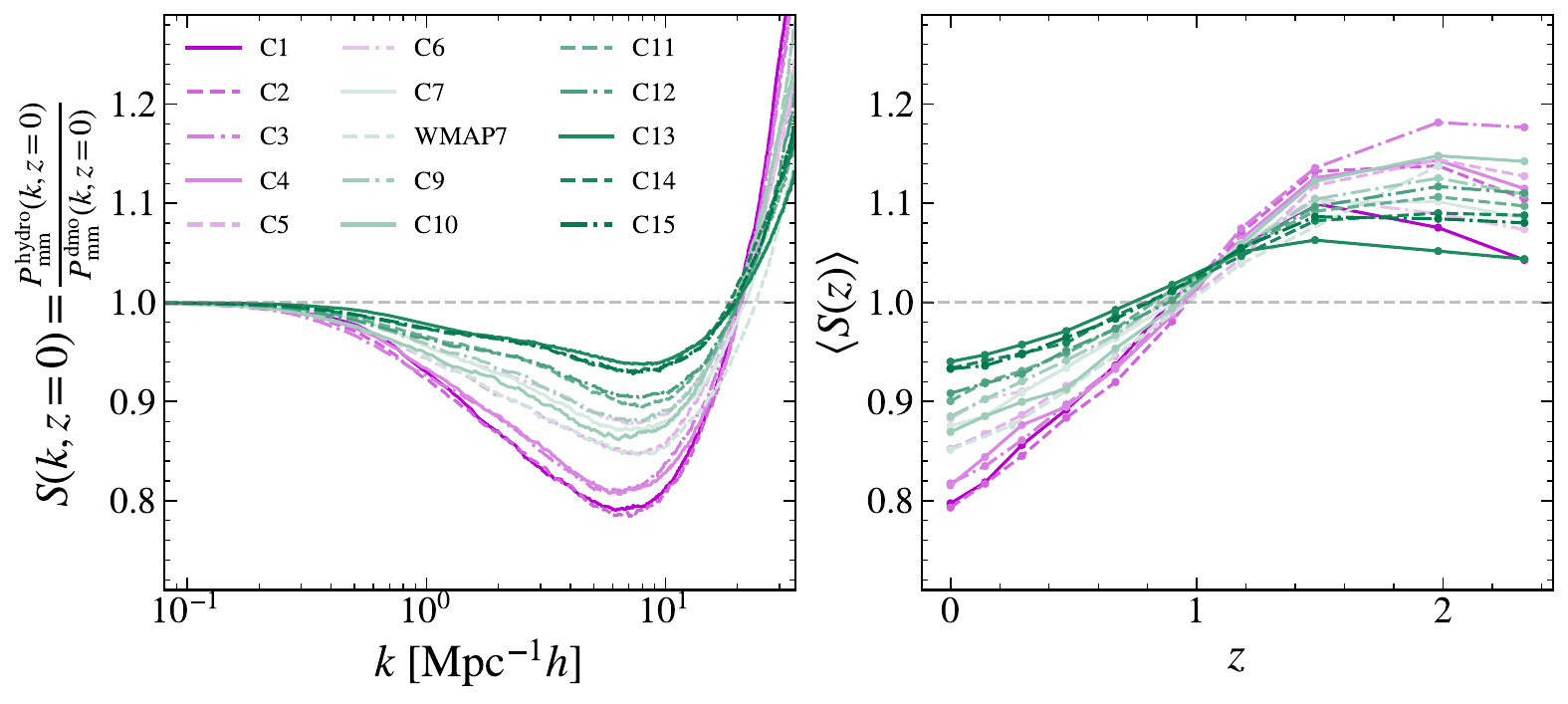}
    \caption{Suppression in the matter power spectrum for the \textit{Magneticum} simulations. Left: Suppression as a function of wavenumber at $z=0$. Right: Suppression averaged between $k=1-10\,\text{Mpc}^{-1}h$ as a function of redshift.}
    \label{fig:Pk_ratio_z0}
\end{figure*}

The background cosmology dictates the hierarchical growth of dark matter halos, within which galaxies form and evolve. Primarily, the balance between the gravitational potential of a halo and the strength of feedback mechanisms determines the efficacy of baryonic feedback in redistributing matter \citep{Cui2014,Velliscig2014}. The resulting suppression in the matter power spectrum reflects the cumulative effect across the halo population. Consequently, variations in dark matter halo properties or the halo population itself, driven by changes in cosmology, can directly influence the effectiveness of feedback in altering the matter distribution.

For example, in less massive halos supernovae feedback is more important than AGN activity, while more massive halos, with their deeper potential wells, are less susceptible to AGN feedback \citep{Cui2014,Velliscig2014, Chua2022}. As a result, changes in the halo abundance at a given mass (e.g. due to a different growth rate) can lead to a different level of suppression in the matter power spectrum. Halo concentration which has been found to depend on cosmology \citep{Kwan2013, Ragagnin2021, Cano2022}, also modulates the gravitational potential and affects the extent to which AGN feedback can expel gas. Furthermore, the cosmic baryon fraction influences the gas content of halos and thus impacts the reservoirs available to fuel AGN activity.

\citep{Elbers2024} explored several such mechanisms using the suppression in halo mass --- the ratio of halo mass in the hydrodynamical and dark matter-only simulation --- as a proxy. Specifically, they investigate the impact of halo concentration, formation epoch, and environment using the FLAMINGO suite of hydrodynamical simulations \citep{Schaye2023}. They find that for massive halos ($>10^{13}\, \text{M}_\odot$), a higher concentration leads to smaller baryonic suppression due to the increased gravitational binding energy whereas in the low mass regime ($<5\times10^{12}\, \text{M}_\odot$) this trend reverses. The trend for low mass halos is a result of concentration being anti-correlated with the formation epoch. Halos that form earlier are more concentrated but also have a larger black hole mass. For low mass halos, the non-linear black hole growth (as opposed to self-regulating black holes found in higher mass halos) makes AGN outflows more effective at expelling gas despite a higher concentration. They do not find any correlation between the suppression in halo mass and halo environment, which implies that the cosmology variations do not regulate feedback through this channel.

\begin{figure}
\begin{floatrow}
\ffigbox{%
\includegraphics[width=\linewidth]{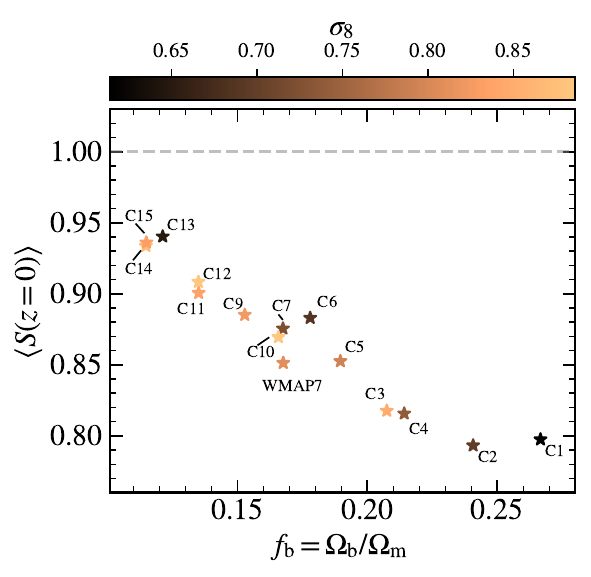}}{\caption{Average suppression in the power spectrum at $z=0$ as a function of the baryon fraction. Overall we see that the suppression is primarily determined by the baryon fraction. C14 and C15 offset vertically for visual clarity.} \label{fig:Pk_fb}}

\ffigbox{%
\includegraphics[width=\linewidth]{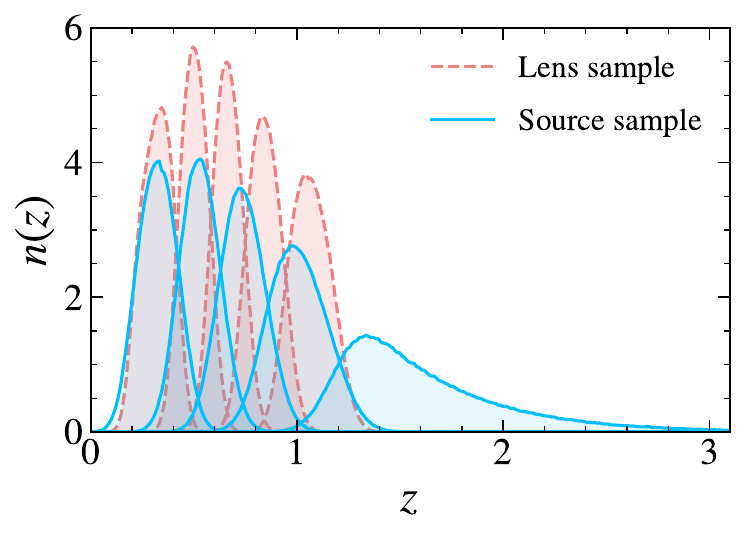}}
{\caption{Normalized redshift distributions for LSST Y1. Refer to section \ref{sec:survey_design} for details.} 
\label{fig:lsst_nz}
}

\end{floatrow}
\end{figure}
To demonstrate the impact of cosmological variations, figure \ref{fig:Pk_ratio_z0} shows the suppression in the matter power spectrum from the \textit{Magneticum} simulations at 15 different cosmologies. The suppression $S(k,z)$ is defined as

\begin{equation} S(k, z) = \frac{P^\mathrm{hydro}_\text{mm}(k, z)}{P^\mathrm{dmo}_\text{mm}(k, z)}, \label{eq:Sk}
\end{equation}
where $P^\mathrm{hydro}_\text{mm}(k)$ and $P^\mathrm{dmo}_\text{mm}(k)$ are the matter power spectra from hydrodynamical and dark matter-only simulations, respectively, for a given cosmology. The results in the left panel show that, even for the same subgrid physics, the strength of baryonic feedback can vary up to 15\% for the cosmology variations considered here. In the right panel we show for each cosmology, the average suppression as a function of redshift, defined as
\begin{equation}
    \langle S(z)\rangle = \frac{\int k^3S(k,z) dk}{\int k^3dk},
\end{equation}
where the lower and upper integration limits are $k=1$ and 10 $\text{Mpc}^{-1}h$, respectively. We see that the evolution of baryonic feedback with redshift varies across cosmologies. For example, cosmologies C3  and C4  exhibit similar levels of suppression at $z = 0$, but differ by approximately 6\% at $z = 2$. Overall, we see that scenarios with stronger suppression also exhibit a stronger evolution with redshift. Accurate modeling of these redshift evolution features might help in breaking  degeneracies between cosmology and baryonic feedback processes.

To explore trends in feedback strength across cosmologies, figure \ref{fig:Pk_fb} presents the average suppression as a function of the cosmic baryon fraction $f_\text{b}=\Omega_\text{b}/\Omega_\text{m}$. We find a strong correlation between suppression in the matter power spectrum and $f_\text{b}$. This correlation can be attributed to dark matter halos having larger gas reservoirs for fueling AGN activity. The strong dependence on $f_b$, even when multiple cosmological parameters are varied simultaneously, implies that the baryon fraction plays a dominant role in influencing feedback strength. These findings are consistent with previous results \citep[see, e.g., Fig. 2 in][]{Schneider2020}, which show that the strength of baryonic effects primarily depends on $f_\text{b}$. We also find that other cosmological parameters, such as $\sigma_8$ and $h$, can have a weak but albeit discernible influence. For example, C1 ($f_\text{b}=0.267$) has a cosmic baryon fraction 10\% larger than C2 ($f_\text{b}=0.241$) and yet displays the same level of suppression. Similarly, C7 and WMAP7 have the same value of $f_\text{b}$ and yet display a 2\% variation in power spectrum suppression.

\section{Simulated likelihood analysis}
\label{sec:analysis}
We perform simulated likelihood analyses using the standard LSST Y1 model implemented in \textsc{Cocoa}\footnote{\href{https://github.com/CosmoLike/cocoa}{\nolinkurl{https://github.com/CosmoLike/cocoa}}}
\citep{Miranda2022}, the latter integrates the theoretical modeling from \textsc{CosmoLike} \citep{cosmolike} and the \textsc{cobaya} framework \citep{cobaya}. We build mock data vectors at different cosmologies by using the power spectrum suppression measured from the \textit{Magneticum} simulations. 
The data vector comprises the 2pt correlation functions of cosmic shear, galaxy-galaxy lensing, and galaxy clustering. 

We now review the modeling ingredients for the simulated analyses including the computation of theory and mock data vectors, baryon mitigation techniques, and survey design.

\subsection{Theory data vector}
\label{sec:theory}
We compute the 2pt correlation functions from the projected angular power spectrum, the latter is computed using the Limber approximation and adopting a flat sky geometry.
For two fields $A$ and $B$, the angular power spectrum in the tomographic bin $(i,j)$ is calculated from the 3D power spectrum $P_{AB}$ as
\begin{equation}
\label{eq:limber}
C_{AB}^{i j}(\ell)= \int_0^{\chi_{\mathrm{h}}} \mathrm{d} \chi \frac{q_A^i(\chi) q_B^j(\chi)}{\chi^2} P_{AB}^{ij}\left(\frac{\ell+1/2}{\chi}, \chi\right),
\end{equation}
where $q^i(\chi)$ is the weight kernel, $\chi$ is the comoving distance, 
 and $\chi_h$ is the comoving horizon distance. The weight kernels for the lensing convergence field $\kappa$ and the galaxy density field $\delta_g$ are given by
\begin{align}
    q_\kappa^{i}(\chi) &= \frac{3 H_0^2 \Omega_m}{2 c^2}\frac{\chi}{a(\chi)}\int_\chi^{\chi_h} \mathrm{d}\chi' \frac{\mathrm{d}z}{\mathrm{d}\chi'} \frac{n^i(\chi')}{\bar{n}^i}\frac{\chi'-\chi}{\chi'},\\
    q^i_{\delta_g}(\chi) &= \frac{n^i(\chi)}{\bar{n}^i} \frac{\mathrm{d}z}{\mathrm{d}\chi},
\end{align}
where $ n^i(\chi'(z))$ is the galaxy redshift distribution  and $a$ is the scale factor. The 3D power spectra for the fields read
\begin{subequations}
\begin{align}
    &P^{ij}_{\kappa\kappa} (k, z)  = P_{\text{mm}}(k, z),\label{eq:P_ss}\\
    &P^{ij}_{\delta_g\kappa} (k, z)  = b^{i}P_{\text{mm}}(k, z),\label{eq:P_gs}\\
    &P^{ij}_{\delta_g\delta_g} (k, z)  = b^{i}b^{j}P_{\text{mm}}(k, z),\label{eq:P_gg}
\end{align}
\end{subequations}
where $b^i$ is the linear galaxy bias and $P_{\text{mm}}(k, z)$ is the non-linear matter power spectrum computed from \textsc{CAMB} \citep{CAMB}. The 3D power spectra in equations \ref{eq:P_ss}, \ref{eq:P_gs} \& \ref{eq:P_gg} are converted to angular power spectra using eq. \ref{eq:limber}, from which the 2pt correlation functions for cosmic shear $\xi_{ \pm}(\theta)$, galaxy-galaxy lensing $\gamma_t(\theta)$, and galaxy clustering $w(\theta)$ are computed as
\begin{align}
\xi_{ \pm}^{i j}(\theta) & =\sum_{\ell} \frac{2 \ell+1}{2 \pi \ell^2(\ell+1)^2}\left[G_{\ell, 2}^{+}(\cos \theta) \pm G_{\ell, 2}^{-}(\cos \theta)\right] C_{\kappa \kappa}^{i j}(\ell), \\
\gamma_t^{i j}(\theta) & =\sum_{\ell} \frac{(2 \ell+1)}{2 \pi \ell^2(\ell+1)^2} P_{\ell}^2(\cos \theta) C_{\delta_g \kappa}^{i j}, \\
w^i(\theta) & =\sum_{\ell} \frac{2 \ell+1}{4 \pi} P_{\ell}(\cos \theta) C_{\delta_g \delta_g}^{i i}(\ell),
\end{align}
where $P_\ell$ and $G_\ell^\pm$ are the Legendre and associated Legendre polynomials, respectively
\citep[cf.][]{Fang_2020}.

\begin{figure*}
    \centering
    \includegraphics[width=\linewidth]{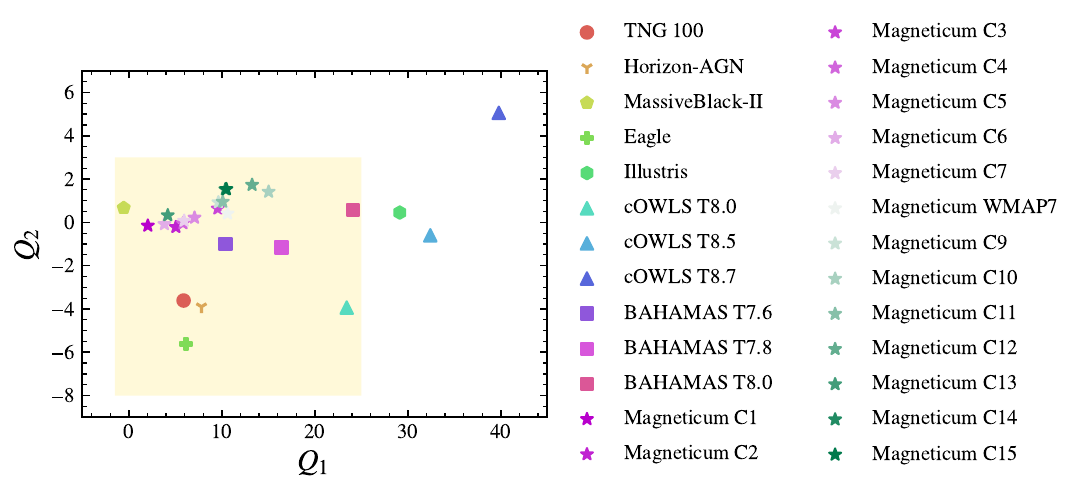}
    \caption{Amplitudes of the first two principal components measured from hydrodynnamical simulations. The amplitudes are estimated by projecting the mock data vector on the PCs, as the latter form an orthogonal basis. The mock data vectors for \textit{Magneticum} simulations are computed at their respective cosmologies while the mock data vector for the remaining simulations assume a \textit{WMAP}7 cosmology. The shaded region represents the informative prior used in the analysis, see Sec. \ref{sec:pca_prior} for details.}
    \label{fig:Q_scatter}
\end{figure*}

\subsection{Systematics}
We include the following systematic effects

\begin{itemize}
\item \textit{Intrinsic alignments}: Intrinsic alignments (IA) contaminate the measured cosmic shear signal by introducing additional correlations between galaxy shapes. We model the IA contribution using a redshift dependent non-linear alignment (NLA) model \citep{Bridle2007} parametrized by the amplitude $a_{\mathrm{IA}}$ and redshift evolution $\eta_{\mathrm{IA}}$
    \begin{equation}
        A_\text{IA}(z) = -a_\text{IA} \frac{C_1\rho_\text{crit}\Omega_\text{m}}{G(z)} \left(\frac{1+z}{1+z_\text{0,IA}}\right)^{\eta_\text{IA}},
    \end{equation}
    where, $C_1\rho_\text{crit}$ = 0.0134, $z_\text{0,IA}$ = 0.62, $\Omega_\text{m}$ is the matter density, and $G(z)$ is the linear growth factor.

\item \textit{Photometric redshifts}: The uncertainty in the redshift distribution  $n^i(z)$ is modeled by including a parameter $\Delta z^i$ for each tomographic bin which shifts the distribution as  
    \begin{align}
        n^i(z) \rightarrow n^i(z + \Delta z^i).
    \end{align}
We use separate shift parameters $\Delta z^i_\text{lens}$ and $\Delta z^i_\text{source}$ for the lens and source samples, respectively.

\item \textit{Shear calibration bias}: Galaxy shapes are biased estimators of the underlying shear and the response of the shear estimator has to be calibrated. The residual biases in the shear calibration in a tomographic bin are quantified by the multiplicative calibration bias $m^i$.

\item \textit{Galaxy bias}: We assume linear galaxy bias when modeling the galaxy-galaxy lensing and galaxy clustering parts of the data vector. In total we have five galaxy bias parameters $b^i$, one for each tomographic bin.
\end{itemize}

\subsection{Baryon mitigation techniques}
We test two methods for modeling baryonic effects. 

\begin{itemize}
\item PCA: The PCA method \citep{Eifler2015, Huang_19} uses the difference between the mock data vectors from simulations and theory predicted data vectors, where each data vector consists of ($\xi_\pm(\theta), \gamma_\text{t}(\theta), w(\theta)$),  to identify principal components. These PCs serve as a flexible basis for modeling baryonic effects on smaller scales. For details about the computation of principal components we refer the reader to \cite{Huang_19}. Note that the \textit{Magneticum} simulations, which are used for computing mock data vectors, do not enter PCA training. The model prediction $\mathbf{D}_\mathrm{model}$ is given as the of sum the theory data vector $\mathbf{D}_\mathrm{theory}$ and the principal components $\mathbf{P}_{i}$ weighted by amplitudes $Q_i$
\begin{align}
\label{eq:d_baryon}
    \mathbf{D}_\mathrm{model} =   \mathbf{D}_\mathrm{theory} + \sum_{i=1}^{N} Q_i \mathbf{P}_{i},
\end{align}
where $N$ is the number of principal components. The theory data vector  is generated using the non-linear matter power spectrum from \textsc{halofit} \citep{Takahashi_2012} following the procedure in section \ref{sec:theory}. The PC amplitudes $Q_i$ are treated as free parameters, which allows us to marginalize over baryonic scenarios during cosmological inference. 

The PCA is trained using a difference matrix that depends on both theoretical and mock data vectors. Consequently, the resulting eigenmodes are sensitive to the cosmology at which these data vectors are computed. To train the PCA, we adopt the \textit{WMAP}7 cosmology \citep{Komatsu2011} as the baseline for two key reasons: (i) approximately half of the simulations used for PCA training (described in section \ref{sec:sim_PCA}) are run at this cosmology, and (ii) the \textit{Magneticum} C8 simulation is also run at this exact cosmology which allows us to evaluate the PCA’s ability to capture the subgrid physics implemented in \textit{Magneticum}.

The galaxy clustering and galaxy-galaxy lensing parts of the data vector are robust to baryonic feedback due to the choice of scale cuts (cf. section \ref{sec:survey_design}). Therefore, the principal components only apply corrections to cosmic shear.

\item \textsc{HMCode2020}: The \textsc{HMCode2020} (hereafter, HM20) framework takes a halo model approach to model the effects of non-gravitational physics on the matter power spectrum the \citep{Mead_2015,Mead_2021}. The model comprises parameters that govern the change in the internal structure of dark matter halos, specifically, the modification in halo concentration ($B$), the formation of stars which dominate the matter profile in the halo centers ($f_*$), and the expulsion of gas from halos ($M_\text{b}$). These parameters are calibrated using the power spectrum response from the BAHAMAS \textit{WMAP}9 simulations for wavenumbers in the range $k=0.3\, \text{to}\, 20\, \text{Mpc}^{-1}h$ at $z<1$. HM20 is able to match the measurements from the simulation upto a few percent in the fitted range. In the commonly used single parameter variant of HM20, the model parameters are expressed as a function of the AGN feedback strength $\Theta_\text{AGN}\equiv\log (T_\text{AGN}/\text{K})$, here, $T_\text{AGN}$ is related to the heating temperature of gas particles when an AGN feedback event occurs.

For generating the model data vector with HM20, we evaluate the equations in section \ref{sec:theory} using the \texttt{mead2020\_feedback} option in \textsc{CAMB}.
\end{itemize}

\subsection{Mock data}
To build mock data vectors at different cosmologies which are \textit{contaminated} by  baryonic effects, we use the power spectrum suppression inferred from the \textit{Magneticum} simulations and modify the theoretical prediction (e.g. from \textsc{halofit}) which does not model these effects. For a simulation at a cosmology $\boldsymbol{\theta}_\mathrm{sim}$, we compute the power suppression $S(k, z)$ as a function of wavenumber $k$ and redshift $z$ by taking the ratio of the matter power spectrum from the hydrodynamical simulation and its dark matter-only counterpart (equation \ref{eq:Sk}). We use $S(k, z|\boldsymbol{\theta}_\mathrm{sim})$ to modify the theory prediction of the non-linear matter power spectrum, $P_\text{mm}$, as 
\begin{equation}
    P_\text{mm}(k,z|\boldsymbol{\theta}_\mathrm{sim}) \rightarrow S(k, z|\boldsymbol{\theta}_\mathrm{sim})\times P_\text{mm}(k,z|\boldsymbol{\theta}_\mathrm{sim}).
\end{equation}
For the PCA method, we compute $P_\text{mm}$ using the \textsc{halofit} option in \textsc{CAMB}. For HM20, however, the \texttt{mead2020\_feedback} model which applies baryonic corrections and is parameterized by $\Theta_\text{AGN}$, does not recover the dark matter-only prediction under any limit of $\Theta_\text{AGN}$. Therefore, we compute the mock data vector for HM20 by modifying $P_\text{mm}$ from the \texttt{mead2020} option. We then proceed to compute mock data using the contaminated matter power spectrum and following the modeling recipe described in section \ref{sec:theory}.

\begin{table}
\centering
\begin{tabular}{|| l l l r ||}
\hline
Parameter & & Prior \hspace{10pt} &   Fiducial Value \\
\hline \multicolumn{4}{||c||}{ \textbf{Cosmology}} \\
$A_{\mathrm{s}}$ & Flat &(5$\times10^{-10}$, 1$\times 10^{-8}$) & Varied \\
$\Omega_{\mathrm{m}}$ & Flat&$(0.05,0.9)$  & Varied \\
$\omega_\mathrm{b}\equiv\Omega_\mathrm{b}h^2$ &  Gaussian&$(\omega_\mathrm{b}^\mathrm{fid},0.0016)$ &  Varied \\
$h$ & Flat&$(0.55,0.91)$ & Varied \\
$n_{\mathrm{s}}$ & Flat&$(0.87,1.07)$ & 0.963 \\
$\Sigma m_\nu$ &  & Fixed & 0.06 eV\\

\hline \multicolumn{4}{||c||}{ \textbf{Photo-$z$ shift}} \\
$\Delta z^i_{\text{lens}}$ & Gaussian & $(0.0,0.005)$ & 0.0 \\
$\Delta z^i_{\text{source}}$ & Gaussian & $(0.0,0.002)$ & 0.0 \\

\hline \multicolumn{4}{||c||}{ \textbf{Linear galaxy bias}} \\
$b^i$ & Flat & $(0.8, 3)$ &  $[1.24, 1.36, 1.47, 1.60, 1.76]$\\

\hline \multicolumn{4}{||c||}{\textbf{Shear calibration}} \\
$m^i$ & Gaussian & $(0.0,0.005)$ & 0.0 \\

\hline \multicolumn{4}{||c||}{\textbf{Intrinsic Alignment}} \\
$A_{\mathrm{IA}}$ & Flat & $(-5,5)$ & 0.5 \\
$\eta_{\mathrm{IA}}$ & Flat & $(-5,5)$ & 0.0 \\

\hline \multicolumn{4}{||c||}{\textbf{Baryon Model: PCA}} \\
$Q_1$ & Flat & $(-1, 25)$ & - \\ 
$Q_2$ & Flat & $(-8, 3)$ & - \\ 

\hline \multicolumn{4}{||c||}{\textbf{Baryon Model: HM20}} \\
$\Theta_\text{AGN}\equiv\log_{10} (T_{\text{AGN}}/$K) & Flat & $(7, 8)$ & - \\
\hline
\end{tabular}
\caption{Model parameters and priors for the simulated likelihood analyses. Flat$(a,b)$ denotes a flat prior within lower and upper limits $a$ and $b$, respectively. Gaussian$(\mu,\sigma)$ denotes a Gaussian prior with mean $\mu$ and standard deviation $\sigma$. The fiducial values for cosmology parameters are enumerated in table \ref{tab:box3_cosmo}.}
\label{tab:model_priors}
\end{table}
\subsection{Survey Design}
\label{sec:survey_design}
We perform a simulated LSST Y1 analysis, following the survey design outlined in the DESC Science Requirements Document (DESC SRD, \citep{DESCSRD2018}). The analysis assumes a survey area of 12,300 deg$^2$, with redshift distributions for the lens and source samples derived using limiting $i$-band magnitudes of 24.1 and 25.1, respectively. The redshift distributions are modeled using the analytical form $n(z)\propto z^2 \text{exp}[-(z/z_0)^\alpha]$. For the lens sample, we set  $(z_0,\alpha)=(0.26,0.94)$ and normalize the distribution by an effective number density $n_\text{eff}=18\,\,\text{arcmin}^{-2}$. The sample is then divided into five equally populated tomographic bins (differing from the DESC SRD, which uses 10 bins) based on estimated redshift and each bin is convolved with a Gaussian photo-$z$ scatter of $\sigma_z= 0.03(1 + z)$. For the source sample, we use $(z_0,\alpha)=(0.19,0.87)$ and normalize the distribution to $n_\text{eff}=11.2\,\,\text{arcmin}^{-2}$. The source sample is also split into five tomographic bins, with each bin convolved with a Gaussian photo-$z$ uncertainty of $\sigma_z= 0.05(1 + z)$. The lens and source redshift distributions are shown in figure \ref{fig:lsst_nz}.

Following the DESC SRD, we choose a scale cut of $k_\text{max}=0.3\,\text{Mpc}^{-1}h$, which approximately corresponds to a minimum comoving scale $R_{\text{min}}=2\pi/k_\text{max}=21\, \text{Mpc}\,h^{-1}$. For galaxy clustering, the angular scale cut for a tomographic bin $i$ is $\theta_{\text{min}}^{i}=R_{\text{min}}/\chi(\Bar{z}^i)$, where $\Bar{z}^i$ is the mean redshift of the tomographic bin. Thus for the lens sample, the minimum comoving scale translates to angular scale cuts of $[105.86',\,  81.17',\,  71.24',\,  65.32',\, 61.49']$; we adopt the same angular scale cuts for galaxy-galaxy lensing. For cosmic shear, we restrict the analysis  to an angular multipoles of $\ell<\ell_\mathrm{max}=5000$, which is more aggressive than the scale cut of $\ell_\mathrm{max}=3000$ specified in the DESC SRD. The angular scale cuts for the correlation functions $\xi_\pm$ are determined by the first zeros of their corresponding spherical Bessel function $J_{0/4}(\ell\theta)$, which translates to a minimum angular scales of $\theta_\mathrm{min}^{\xi_+}=2.4048/\ell_\mathrm{max}=1.653'$ and $\theta_\mathrm{min}^{\xi_-}=7.5883/\ell_\mathrm{max}=5.217'$.

Note that as a consequence of the scale cuts described above, the galaxy clustering and galaxy-galaxy lensing data vectors are robust to baryonic effects and thus the constraints on baryonic feedback parameters are driven by cosmic shear data only.

\subsection{Likelihood and covariance}
We use the Python package \textsc{emcee} \citep{emcee} to sample the posterior distribution assuming a Gaussian likelihood
\begin{align}
    \log \mathcal{L} = -\frac{1}{2}(\mathbf{D}_\mathrm{mock}-\mathbf{D}_\mathrm{model})^\mathrm{T} \mathbf{C}^{-1}(\mathbf{D}_\mathrm{mock}-\mathbf{D}_\mathrm{model}),
\end{align}
where $\mathbf{C}$ and $\mathbf{D}$ refer to the covariance and data vector, respectively, and the subscript denotes if the latter is the mock or the model data vector. We use a Gaussian prior on the physical baryon density $\Omega_\text{b}h^2$ with the prior width set to ten times the measurement uncertainty from \textit{Planck} \citep{Planck2020}.
The model parameters priors are summarized in Table \ref{tab:model_priors}. 

The covariance for the 3$\times$2pt data vector is calculated using the publicly available code \textsc{CosmoCov}\footnote{\href{https://github.com/CosmoLike/CosmoCov}{\nolinkurl{https://github.com/CosmoLike/CosmoCov}}} \citep{cosmocov} which is built upon the \textsc{CosmoLike} framework. \textsc{CosmoCov} uses the 2D-FFTLog algorithm to efficiently compute real-space covariance matrices. We compute the cosmic shear covariance which consists of Gaussian, super-sample covariance and connected non-Gaussian components.

Since the model data vector evaluations are slow and computationally expensive, we use the neural network emulator presented in \citep{Boruah2023} which predicts the model data vector as function of cosmology and nuisance parameters. We train separate emulators for each component of the data vector ($\xi_+$, $\xi_-$, $\gamma_t$, $w$). As the fiducial cosmologies for the \textit{Magneticum} simulations differ significantly, we train separate emulators for each cosmology to ensure prediction accuracy. We refer the reader to \citep{Boruah2023} for further details about the emulator architecture and training method.

\subsubsection{Prior on baryonic feedback parameters}
\label{sec:pca_prior}
The simulations used for constructing the PC basis can serve as external information to inform priors that better match observations.
Figure \ref{fig:Q_scatter} shows the distribution of the PC amplitudes $\{Q_i \}$ for the simulations considered in this work\footnote{The amplitudes for each simulations are estimated by projecting the difference between the mock and theory data vectors onto the principal eigenmodes. Since the PCs form an orthogonal basis, by construction, the amplitude is simply $Q_i=\frac{1}{||\mathbf{P}_i||}\mathbf{P}_i^\text{T}(\mathbf{D}_\text{mock}-\mathbf{D}_\text{theory})$}. We derive informative priors for the PC amplitudes by examining the range  of $\{Q_i \}$ spanned in the figures (not including the \textit{Magneticum} simulations), as indicated by the shaded region. We exclude the Illustris, cOWLS T8.5, and cOWLS T8.7 simulations when determining these priors as they represent too extreme feedback scenarios. The baryon fractions in group and cluster sized halos predicted by the Illustris simulation are in disagreement with observations due to the strong AGN radio-mode feedback \citep{Genel2014, Haider2016}. The cOWLS T8.5 and cOWLS T8.7 simulations also do not reproduce the observed gas fractions in massive halos \citep{LeBrun2014}.

\begin{figure}
    \centering
    \includegraphics[width=\linewidth]{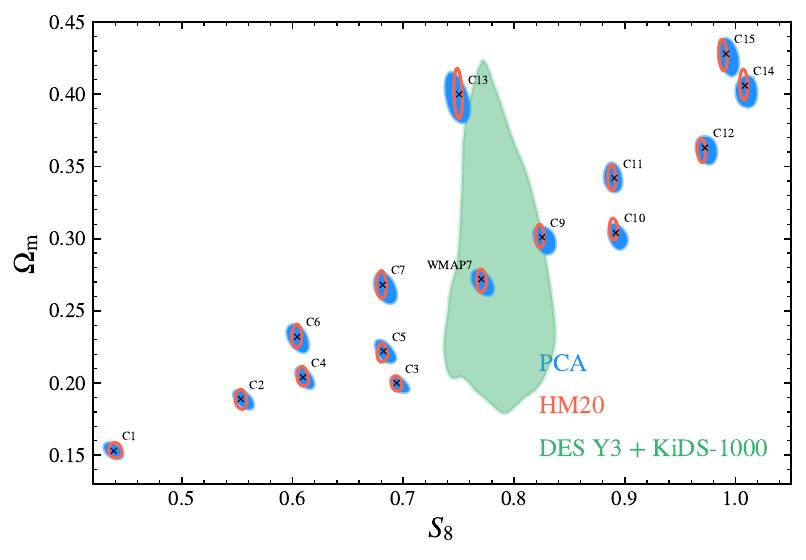}
    \caption{Results from simulated 3$\times$2pt analysis  as a function of cosmology. The figure shows  1$\sigma$ contours (39.35\% credible intervals)  in $\Omega_\text{m}$--$S_8$ when a mock data vector contaminated by baryonic effects is fit using PCA and HM20. At each cosmology, we compute the mock data vector by modifying the non-linear matter power spectrum by the suppression measured from the \textit{Magneticum} simulation. The cross marker represents the true values (refer to table \ref{tab:box3_cosmo} for list of parameter values at the 15 cosmologies). For reference we also show 3$\sigma$ contours (98.9\% credible intervals) from the joint DESY3 + KiDS 1000 analysis \citep{DESY3_KIDS}. Note that the constraints shown in this figure include projection effects, however, they're insignificant at the scale of this visualization.}
    \label{fig:MC_contours}
\end{figure}

\section{Results}
\label{sec:results}

\subsection{Performance at the baseline \textit{WMAP}7 cosmology}
We first asses the performance of the baryon mitigation techniques in capturing the  baryonic feedback in the \textit{Magneticum} \textit{WMAP}7 (C8) simulation. Note that, while the principal components are computed at the \textit{WMAP}7 cosmology, the \textit{Magneticum} simulations are not used for training the PCA. Similarly, HM20 is calibrated on the BAHAMAS \textit{WMAP9} simulation which differs from \textit{Magneticum} in its subgrid physics implementation. Therefore, it is important to first verify if both methods result in unbiased inference at this baseline scenario before considering variations in cosmology.

For a quantitative assessment of the performance of the two methods, we adopt the bias in the $\Omega_\mathrm{m}$--$S_8$ plane as our metric, as the two parameters are of key interest for weak lensing surveys. We consider the baryon model to be effective if the bias in the 2D  marginalized mean is below 0.3$\sigma$. The bias in mean parameter values can also result from the marginalization in the high dimensional parameter space. To account for these so-called projection effects, we repeat the analysis by fitting a data vector generated from the model and adopt the resulting 2D marginalized mean as the reference, rather than the fiducial parameter values, for computing the bias\footnote{The bias is computed as $\frac{\text{bias}}{\sigma}=\sqrt{\textbf{X}^\text{T}\textbf{C}_{\Omega_\mathrm{m}-S_8}^{-1}\textbf{X}}$, where $\textbf{X}=(\Delta\Omega_\mathrm{m}, \Delta S_8)^\text{T}$ is the difference between the 2D marginalised mean and the \textit{reference} parameter values. $\textbf{C}_{\Omega_\mathrm{m}-S_8}$ is the marginalized parameter covariance in the $\Omega_\mathrm{m}$--$S_8$ plane.}. Using the same model for, both, generating and fitting the data vector eliminates model misspecification and isolates the bias due to projection effects, enabling us separate it from the systematic effect of interest.

The two baryon models exhibit comparable accuracy in the 3$\times$2pt analysis at the baseline cosmology. For PCA and HM20, we recover cosmological parameters within $0.1\sigma$ and $0.12\sigma$, respectively, indicating that both methods effectively capture the subgrid physics implementation in \textit{Magneticum}. Therefore, any bias observed when repeating the analysis for other cosmologies would therefore arise from variations in cosmology and the resulting modulation of suppression in the matter power spectrum.

\begin{figure}
    \centering
  \includegraphics[width=\linewidth]{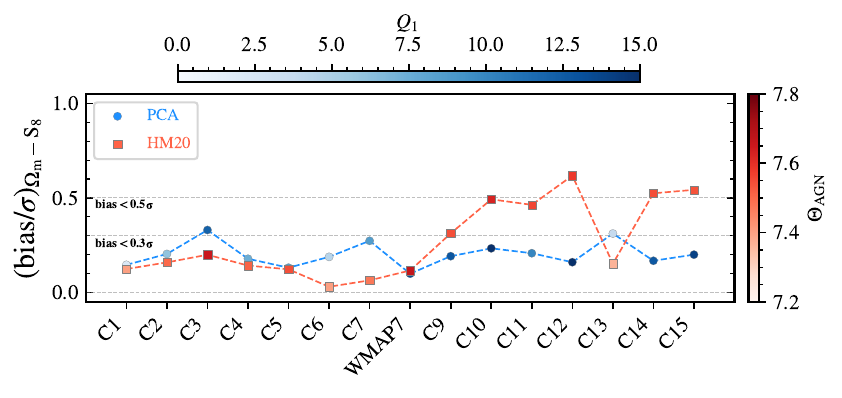}
    \caption{Bias in $\Omega_\mathrm{m}$--$S_8$ at different cosmologies when using PCA and HM20 to mitigate baryonic effects. The marker color represents the amplitude of baryonic feedback parameter in the respective models, $Q_1$ for PCA and $\Theta_\text{AGN}$ for HM20. The bias when using PCA is $<0.3\sigma\, (0.5\sigma)$ in 13 (all) scenarios indicating that that PCA is robust to the coupling between baryonic feedback and cosmology, meanwhile, using HM20 results in $<0.3\sigma\, (0.5\sigma)$ bias for 9 (12) scenarios. The bias shown here is corrected for projection effects.}
    \label{fig:MC_bias}
\end{figure}

\subsection{Baryon mitigation across cosmologies}
\label{sec:result_MC}

After validating both methods at the baseline cosmology, we perform simulated analysis at the remaining cosmological scenarios. In figure \ref{fig:MC_contours}, we show the  $\Omega_\mathrm{m}$--$S_8$ constraints across the 15 cosmologies using PCA in blue and HM20 in red. The estimated bias at each cosmology is shown in figure \ref{fig:MC_bias}, where we also color the symbol based on baryonic feedback parameter i.e. $Q_1$ for PCA and $\Theta_\text{AGN}$ for HM20. We find that the PCA method obtains unbiased parameter constraints in all but two scenarios. The parameter bias is at most weakly correlated with feedback strength, as measured by the first PC amplitude\footnote{While $Q_1$ represents the level of suppression in the data vector, the values at different cosmologies must be interpreted with caution. This is because $Q_1$ is proportional to the absolute change in the data vector, as evident from equation \ref{eq:d_baryon}; therefore cosmologies with larger $\Omega_\text{m}$ or $\sigma_8$ (and hence larger $P_\text{mm}(k)$ amplitude) will naturally have larger $Q_1$ even if the suppression is relatively small.}. While relaxing the priors on PC amplitudes results in a loss of constraining power, it also makes it more robust to such effects. When using flat wide priors of $(-50,50)$, the bias is $<0.3\sigma$ in all cases. The HM20 model results in biased constraints at six out of the fifteen cosmologies and in three scenarios the bias is $>0.5\sigma$. Note that, neither PCA or HM20 are trained/calibrated on the \textit{Magneticum} simulations

\begin{figure}
    \centering
    \includegraphics[width=0.48\linewidth]{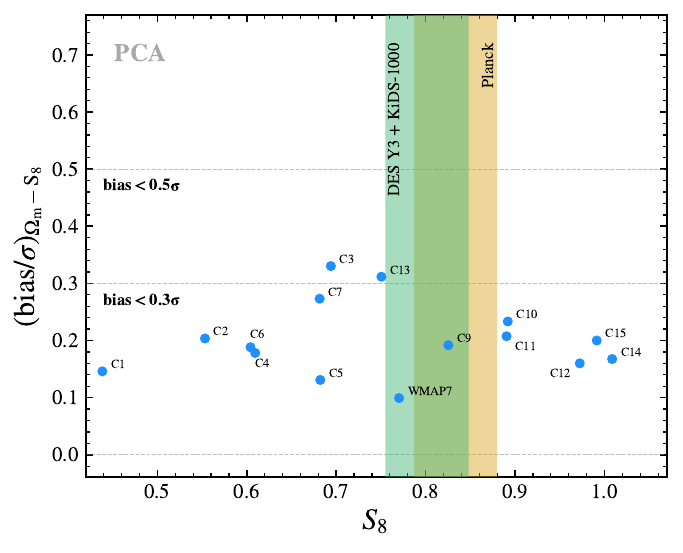}
    \includegraphics[width=0.48\linewidth]{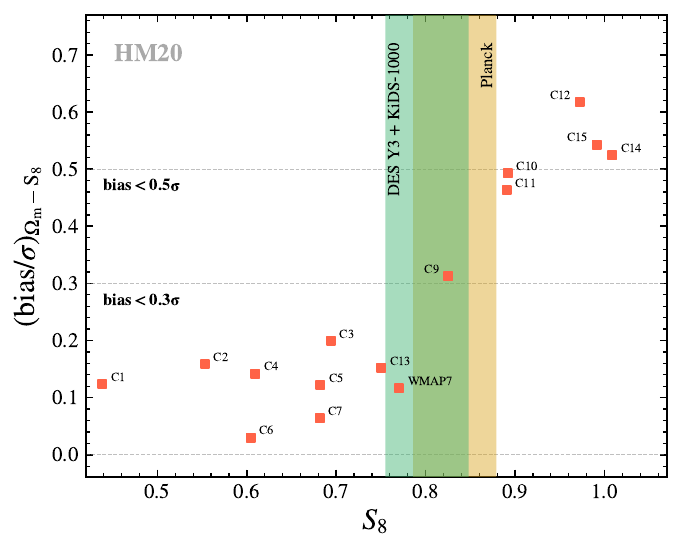}
    \caption{Bias in $\Omega_\mathrm{m}$--$S_8$ as a function of $S_8$. Results for PCA and HM20 model are shown in the left and right panels, respectively. The shaded regions represent $3\sigma$ (99.7\% confidence levels) constraints from DES Y3 + KiDS-1000 analysis \citep{DESY3_KIDS} and \textit{Planck} \citep{Planck2020}. We do not find a significant trends for PCA, while the bias for HM20 shows strong correlation with the input cosmology. Note that the parameter bias shown here is corrected for projection effects.}
    \label{fig:bias_hmcode}
\end{figure}

The bias when using the HM20 model correlates with the input cosmological parameters (most notably $S_8$) while we do not observe any such trend for PCA, as shown in figure \ref{fig:bias_hmcode}. For comparison, we show 3$\sigma$ constraints from the DES Y3 + KiDS-1000 analysis \citep{DESY3_KIDS} and \textit{Planck} \citep{Planck2020} as shaded regions. Even when considering only the cosmologies that are feasible given the constraining power of existing surveys, these results suggests that the bias induced due to a coupling with cosmology can still be significant for HM20. This implies that HM20 may not fully capture the impact of cosmology on feedback strength at LSST Y1 accuracy. This is corroborated by previous findings that the HM20 model calibrated using the BAHAMAS \textit{WMAP}9 simulation shows differences of a few percent when it is used to predict the suppression in BAHAMAS \textit{Planck}2013 simulation \citep[see Sec. 6.3 in][]{Mead_2021}. 
Another possibility is that the HM20 predictions get progressively inaccurate at higher redshifts as the model is calibrated only at $z<1$.

\begin{figure}
    \centering
    \includegraphics[width=\linewidth]{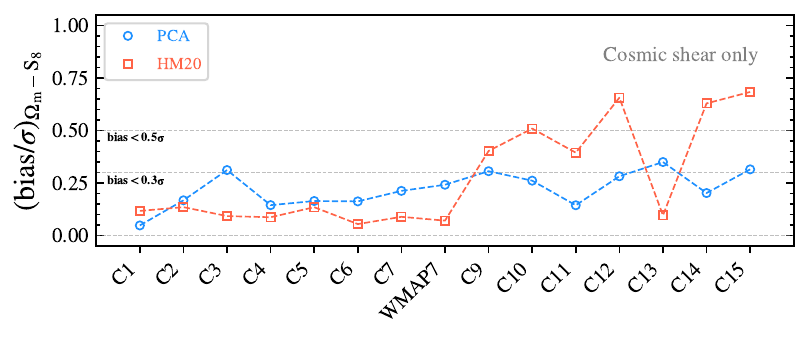}
    \caption{Bias in $\Omega_\mathrm{m}$--$S_8$ when analyzing a cosmic shear only data vector, corrected for projection effects. The bias for both baryon mitigation methods is larger compared to the 3$\times$2pt case (figure \ref{fig:MC_bias}).}
    \label{fig:bias_cs}
\end{figure}
One way to enhance the performance of HM20 is to introduce greater model flexibility. Instead of parameterizing in terms of the subgrid heating temperature, $\Theta_\text{AGN}$, we can directly fit the model parameters ($B, f_*, M_\text{b}$) along with their redshift dependence. While this approach may introduce parameter degeneracies, applying informative priors based on simulations can mitigate this issue. For example, \citep{Mead_2021} provides priors derived from fits to the BAHAMAS and cosmo-OWLS simulations. Additionally, exploring alternative parameterizations of redshift evolution could improve accuracy, as \citep{Mead_2021} found that the power-law $z$-dependence adopted in HM20 led to inaccurate fits for $z>1$. The performance of PCA is also dependent on the simulation set used for training and the prior on PC amplitudes, therefore the precision is likely to improve as more realistic simulations are included in the training set.

We repeat the simulated analyses using only the cosmic shear component of the data vector and observe a degradation in the performance for both PCA and HM20, as shown in figure \ref{fig:bias_cs}. While the galaxy clustering and galaxy-galaxy lensing portions of our data vector are insensitive to baryonic feedback due to the scale cuts applied, including probes such as the Sunyaev-Zeldovich effects, which incorporate both cosmological and astrophysical information, can help disentangle parameter degeneracies and mitigate the impact of the coupling \citep{Fang2024}.

\section{Conclusions}
\label{sec:conclusions}
Stage IV surveys will deliver vast volumes of data, enabling an unprecedented level of precision in cosmological measurements and inference. To fully leverage the statistical power of these datasets, these advances in data acquisition must be matched by improvements in modeling techniques. Of particular importance is understanding the growth of structure on non-linear scales. In the context of weak lensing analyses, the modeling uncertainty on small scales is dominated by baryonic processes like stellar and AGN feedback.

Current analyses generally assume the impact of baryonic feedback on the matter distribution to be independent of the underlying cosmology. We test this assumption and quantify its impact on parameter inference using the \textit{Magneticum} suite of hydrodynamical simulations. These simulations are run at a fixed calibration of subgrid physics parameters which allows us to isolate the dependence of baryonic feedback on cosmology. We find that the suppression in the matter power varies up to 15\% across the cosmologies spanned by the simulations. We show that the suppression in the power spectrum is primarily determined by the cosmic baryon fraction $f_\text{b}=\Omega_\text{b}/\Omega_\text{m}$ with other cosmological parameters resulting in a few percent effect at most.

We generate mock 3$\times$2pt data vectors by modifying the non-linear matter power spectrum with the suppression measured from the 15 \textit{Magneticum} simulations. We test two baryon mitigation techniques, PCA and HM20, by performing a simulated analyses for a LSST Y1 like survey. Our analysis shows that the PCA method is  robust to cosmological variations, while HM20 exhibits a bias greater than $0.3\sigma$ in $\Omega_\text{m}$--$S_8$ at six out of fifteen cosmologies.

For the PCA method, we do not observe significant correlation between the bias and either feedback strength or the fiducial cosmology, suggesting that PCA effectively captures these features. We find that the bias in parameter inference from the HM20 model is correlated with the background cosmology. This indicates that the dependence on cosmology assumed in HM20 (through the cosmic baryon fraction) might not be sufficiently accurate and needs further investigation. As the redshift evolution of feedback strength also changes with cosmology, the performance of HM20 might also be limited by the redshift evolution of parameters assumed in the model. To improve model flexibility, one approach is to relax current assumptions by introducing additional parameters and/or allowing existing parameters to vary independently, rather than being tied to $\Theta_\text{AGN}$. Advancements in hydrodynamical simulations will be crucial for enhancing the performance of both baryon mitigation techniques. Simulations suites such as ANTILLES \citep{Salcido2023} and FLAMINGO \citep{Schaye2023} span wider range of possible feedback scenarios and can thus be used to determine realistic priors for model parameters.

In conclusion, cosmological inference at the precision required for upcoming surveys will demand a thorough evaluation of baryon modeling. It is essential to rigorously test these models using mock datasets from hydrodynamical simulations. Since the sensitivity to the coupling between cosmology and baryonic feedback is likely influenced by subgrid physics parameters, the performance of baryon models must be assessed across data vectors that span a range of cosmologies and feedback implementations.

% \newpage
\appendix

\acknowledgments
We thank Tiago Castro for assistance with the Magneticum measurements and insightful discussions during the initial stages of this work.
PRS also thanks Shivam Pandey for helpful discussions. PRS is supported in part by a Joint Fellowship Program between the University of Arizona and CNRS. PRS and EK are supported in part by Department of Energy grant DE-SC0020247 and the David and Lucile Packard Foundation. KD acknowledges support by the COMPLEX project from the European Research Council (ERC) under the European Union’s Horizon 2020 research and innovation program grant agreement ERC-2019-AdG 882679 as well as
 by the Deutsche Forschungsgemeinschaft (DFG, German Research Foundation) under Germany’s Excellence Strategy - EXC-2094 - 390783311. The calculations for the hydrodynamical simulations were carried out at the Leibniz Supercomputer Center (LRZ) under the project pr83li (Magneticum). KB, EA, and YD  received funding from the Centre National d’Etudes Spatiales for the completion of this work. The analyses in this work were carried out using the High Performance Computing (HPC) resources supported by the University of Arizona Technology and Research Initiative Fund (TRIF), University Information Technology Services (UITS), and the office for Research, Innovation, and Impact (RDI) and maintained by the UA Research Technologies Department.

% Bibliography

%% [A] Recommended: using JHEP.bst file
\bibliography{biblio.bib}
\bibliographystyle{JHEP}

\end{document}